\documentclass[prl,twocolumn,superscriptaddress,showpacs]{revtex4}
\usepackage{amsmath}
\usepackage{amsfonts}
\usepackage{amssymb}
\usepackage{graphicx}

\begin{document}

\title{Complementarity and Information in "Delayed-choice for entanglement swapping"\footnote{This
paper is dedicated to Prof. Asher Peres on the occasion of his
70th birthday.}}

\author{{\v C}aslav Brukner\footnote{Present address: Optics Section, The
Blackett Laboratory, Imperial College, Prince Consort Road,
London, SW7 2BW, United Kingdom}}
\email{caslav.brukner@univie.ac.at} \affiliation{Institut f\"ur
Experimentalphysik, Universit\"at Wien, Boltzmanngasse 5, A--1090
Wien, Austria}
\author{Markus Aspelmeyer}
\email{markus.aspelmeyer@univie.ac.at} \affiliation{Institut f\"ur
Experimentalphysik, Universit\"at Wien, Boltzmanngasse 5, A--1090
Wien, Austria}
\author{Anton Zeilinger}
\email{anton.zeilinger@univie.ac.at} \affiliation{Institut f\"ur
Experimentalphysik, Universit\"at Wien, Boltzmanngasse 5, A--1090
Wien, Austria} \affiliation{Institut f\"ur Quantenoptik und
Quanteninformation, \"Osterreichische Akademie der
Wissenschaften, Boltzmanngasse 3, A--1090 Wien, Austria}

\date{\today}

\begin{abstract}

Building on Peres's idea of ``Delayed-choice for extanglement
swapping" we show that even the degree to which quantum systems
were entangled can be defined after they have been registered and
may even not exist any more. This does not arise as a paradox if
the quantum state is viewed as just a representative of
information. Moreover such a view gives a natural quantification
of the complementarity between the measure of information about
the input state for teleportation and the amount of entanglement
of the resulting swapped entangled state.

\end{abstract}

 \maketitle

Entangled systems display one of the most interesting features of
quantum mechanics - their joint state is not separable regardless
of their spatial separation. It is still sometimes believed that
for obtaining entangled states quantum systems
\textit{necessarily} need to interact (dynamically) with one
another, either directly, or indirectly via other particles.

Yet, an alternative possibility to obtain entanglement is to make
use of projection of the state of two particles onto an entangled
state. The procedure is known as "entanglement swapping" and was
suggested in 1993~\cite{marek} and experimentally demonstrated for
the first time in 1998~\cite{pan}. In the experiment two pairs of
entangled photons $0-1$ and $2-3$ are produced and one photon from
each of the pairs is sent to two separated observers, say photon
$0$ is sent to Alice and photon $3$ to Bob, as schematically shown
in Fig.~\ref{swapping}. The other photons, $1$ from the first pair
and $2$ from the second pair, are sent to the third observer,
Victor. He subjects photons $1$ and $2$ to a Bell-state
measurement, by which photons $0$ and $3$ become automatically
entangled. This requires the entangled photons 0 and 3 neither to
come from a common source nor to have interacted in the past.

A seemingly paradoxical situation arises, if - as suggested by
Peres~\cite{peres} in his paper entitled {\emph{``Delayed-choice
for entanglement swapping''} - ``entanglement is produced \emph{a
posteriori}, after the entangled particles have been measured and
may even no longer exist''. In such a situation particles 0 and 3
are detected before the Bell-state measurement has been performed.
This seems paradoxical, because Victor's measurement projects
photons 0 and 3 into an entangled state \emph{after they have been
registered.} Furthermore, Victor is even free to choose the kind
of measurement he wants to perform on photons $1$ and $2$. Instead
of a Bell-measurement he could also measure the polarization of
these photons individually which would result in a well defined
polarization for photons 0 and 3, i.e. a separable product state.
Therefore, depending on Victor's later measurement, Alice's and
Bob's earlier results indicate that photons $0$ and $3$ were
either entangled or not.

\begin{figure}
\centering
\includegraphics[angle=0,width=7.5cm]{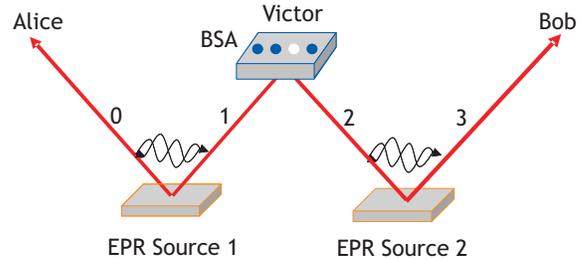}
\caption{Scheme of entanglement swapping. Two pairs of entangled
particles $0-1$ and $2-3$ are produced by two
Einstein-Podolsky-Rosen (EPR) sources. One particle from each of
the pairs is sent to two separated observers, say particle $0$ is
sent to Alice and particle $3$ to Bob. The other particles $1$ and
$2$ from each pair are sent to Victor who subjects them to a
Bell-state analyser (BSA), by which particles $0$ and $3$ become
entangled although they have never interacted in the past.}
\label{swapping}
\end{figure}

In this article we put this paradoxical situation to its extreme
to show that even the \emph{degree} to which the particles were
entangled can be defined after the particles have been registered.
This is because Victor could choose a measurement for photons 1
and 2 projecting them into an \emph{arbitrarily partially}
entangled state after photons 0 and 3 have been detected.
Therefore, depending on the degree of entanglement of the states
into which photons 1 and 2 are projected in Victor's later
measurement, Alice's and Bob's earlier results correspond to a
specific degree of entanglement of photons 0 and 3. Furthermore,
this degree of entanglement is specified by the amount of
information about the state of photon 1, which is gained in
Victor's measurement of photons 1 and 2. Using an
information-theoretic description we give a quantitative
\emph{complementarity relation} for the degree of entanglement of
photons 0 and 3 and the amount of information about the state of
photon 1.

Recently, delayed-choice entanglement swapping in the spirit of
the proposal by Peres~\cite{peres} was experimentally demonstrated
using polarization-entangled photons~\cite{tomas}. In such
experiments, a Bell-state measurement is realized by overlapping
two photons on a balanced (50:50) beamsplitter and analyzing their
distribution in the output ports. Specifically, for a projection
on the spin-singlet state $|\psi^-\rangle$ the two photons exit in
different ports of the beamsplitter. By including two $10m$
optical fibre delays for both outputs of the Bell-state
measurement photons 1 and 2 hit the detectors delayed by about
$50ns$ with respect to detection of photons 0 and 3
(Fig.~\ref{delayed}). Importantly, it was shown that the observed
fidelity of the entangled state of photons 0 and 3 matches the
fidelity in the nondelayed case within experimental errors. This
indicates that relative temporal order of Alice's and Bob's events
on one hand, and Victor's events on the other hand is irrelevant.
It also strengthens the view that what here matters is bringing
lists of Alice's, Bob's and Victor's measurement results into an
appropriate \emph{relation}~\cite{tomas,ryff,replay}. Alice and
Bob can divide their own lists of locally completely random
measurement results into many different subsets of results that
correspond to different possible Victor's measurements. Which
events fall into which subset they might learn in the future from
Victors's measurement results. And it is independent of whether
Victor's measurement has been performed before or even after
Alice's and Bob's measurements. Because the quantum state is no
more than the most compact representative of such
lists~\cite{schroedinger}, it is conceptually allowed to say that
particles can become entangled even after they already have been
registered.

\begin{figure}
\centering
\includegraphics[angle=0,width=5.7cm]{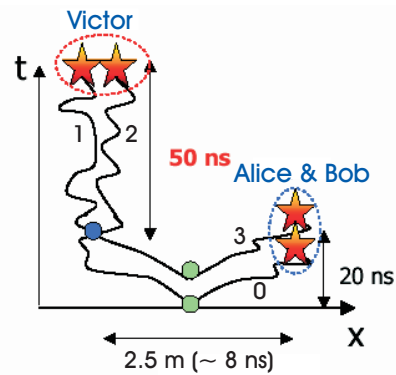}
\caption{Space-time diagram for delayed choice entanglement
swapping experiment reported in Ref.~\cite{tomas}. Alice's and
Bob's detector were located next to each other. The time travel of
photons 0 and 3 from the source to these detectors was equal to
about 20 ns. By including additional optical fiber delays for both
outputs of the Bell-state analyser photons 1 and 2 hit Victor's
detectors delayed by about 50 ns. Victor and the pair Alice and
Bob were separated by about 2.5 m, corresponding to luminal
traveling time of 8 ns between them.} \label{delayed}
\end{figure}

In the present work we will consider entanglement swapping
implemented in terms of states of spin-1/2 particles. Hence the
particle pairs are prepared in one of the four Bell states, e.g.,
in the singlet state
\begin{equation}
|\psi^-\rangle_{01} = \frac{1}{\sqrt{2}} (|z+\rangle_0|z-\rangle_1
- |z-\rangle_0 |z+\rangle_1), \label{singlet}
\end{equation}
where, e.g., $|z+\rangle_0$ describes the state "spin up" of
particle $0$ along $z$-direction. This entangled state contains no
information on the spin of the individual particles; it only
indicates that the two particles will give the opposite results if
their spins are measured along the same directions. Thus, as soon
as a measurement on particle 0 projects it, say, onto state
$|z+\rangle_0$ the state of the other one is determined to be
$|z-\rangle_0$, and vice versa. Thus the correlations for
measurements of spin of the two particles along $z$ can be
represented by the proposition:

(i) \textit{``The two particles have the opposite spin along the
$z$-axis''.}

Because the state (\ref{singlet}) is invariant under identical
local unitary transformations the correlations for spin
measurements along $x$ can be represented by the similar
proposition:

(ii) \textit{``The two particles have the opposite spin along the
$x$-axis''.}

The two propositions fully define the entangled state
(\ref{singlet}). This can be seen as a consequence that two bits
of information are available to define a two-qubit
state~\cite{anton}. The truth value of the proposition
\textit{``The two particles have the opposite spin along the
$y$-axis''}, follows from the truth values of the propositions (i)
and (ii). This is because a joint eigenstate of spin products
$\sigma^1_x \sigma^2_x$ and $\sigma^1_y \sigma^2_y$ is also the
eigenstate of $\sigma^1_z \sigma^2_z= - (\sigma^1_x\sigma^2_x)
(\sigma^1_y\sigma^2_y)$. Here, e.g., $\sigma^1_x$ is spin along
$x$ of particle $1$. In general, the four Bell states can be seen
as representation of the four possible two-bit combinations
(true-true, true-false, false-true, false-false) of the truth
values of the propositions (i) and (ii)~\cite{anton1}.

Initially, in the entanglement swapping experiment, the system is
composed of two independent entangled pairs in the overall state:
\begin{equation}
|\psi_{total}\rangle = |\psi^-\rangle_{01} |\psi^-\rangle_{23}.
\label{initial}
\end{equation}
If Victor now subjects particles 1 and 2 to a measurement in a
Bell-state analyzer, and if he finds them in the state
$|\psi^-\rangle_{12}$, then particles 0 and 3 measured by Alice
and Bob, respectively, will be also in the maximally entangled
state $|\psi^-\rangle_{03}$. The reason for this can be seen using
the logic of propositions. Because we observe particles 1 and 2 in
the state $|\psi^-\rangle_{12}$ we know that whatever the result
measured on particle 1 is, the one measured on particle 2 must be
opposite, if they are measured along the same but otherwise
arbitrary directions. But we had initially prepared the pair of
particles 0 and 1 in the state $|\psi^-\rangle_{01}$ and the pair
2 and 3 in the state $|\psi^-\rangle_{23}$ which means that for
measurements along the same directions particles in both the pair
$0-1$ and in the pair $2-3$ will give opposite results. This is
only possible if results measured on particles 0 and 3 will be
opposite with respect to any chosen measurement basis, which is
the actual symmetry of the spin singlet state. This implies that
the final state of particles 0 and 3 is $|\psi^-\rangle_{03}$.

We now suppose that Victor chooses a measurement for particles 1
and 2 projecting them into one of the following four (orthogonal)
states
\begin{eqnarray}
|\psi^+_\alpha\rangle_{12}&=& \alpha |z+\rangle_1|z-\rangle_2 +
\beta |z-\rangle_1|z+\rangle_2 \label{partialbell1}\\
|\psi^-_\alpha\rangle_{12}&=& \beta |z+\rangle_1|z-\rangle_2 -
\alpha
|z-\rangle_1|z+\rangle_2 \label{bas} \\
|\phi^+_\alpha\rangle_{12}&=& \alpha |z+\rangle_1|z+\rangle_2 +
\beta
|z-\rangle_1|z-\rangle_2 \\
|\phi^-_\alpha\rangle_{12}&=& \beta |z+\rangle_1|z+\rangle_2 -
\alpha |z-\rangle_1|z-\rangle_2 \label{partialbell2}
\end{eqnarray}
with \emph{arbitrary} value of $\alpha$ such that $\alpha^2 +
\beta^2 =1$. For simplicity we assume that $\alpha, \beta \in
\mathbb{R}$. For the cases $\alpha=\beta=\frac{1}{\sqrt{2}}$
($\alpha =1$ or $0$), this projection reduces to the previous
Bell-state measurement (to the measurement of spins along $z$ of
particle 1 and 2 individually).

Including Eq. (\ref{singlet}) and ({\ref{initial}) and rearranging
the resulting terms by expressing states of particles 1 and 2 in
the basis of the states (\ref{partialbell1}-\ref{partialbell2})
leads to
\begin{eqnarray}
|\psi_{total}\rangle &=& \frac{1}{2} [|\psi^+_\alpha\rangle_{03}
|\psi^+_\beta\rangle_{12} - |\psi^-_\alpha\rangle_{03}
|\psi^-_\beta\rangle_{12} \nonumber \\
& &  + |\phi^+_\alpha\rangle_{03} |\phi^+_\beta\rangle_{12} -
|\phi^-_\alpha\rangle_{03} |\phi^-_\beta\rangle_{12}].
\label{entswap}
\end{eqnarray}
Each of the four states (\ref{partialbell1}-\ref{partialbell2})
show either perfect correlations or perfect anticorrelations for
measurements of spins along $z$. Thus each of them can be
described by a definite truth value of proposition (i). Yet, they
are only partially entangled and we cannot assert them a definite
truth value also for proposition (ii). Instead, for example, the
representation of the state $|\psi^-_{\alpha} \rangle_{12}$ in the
eigenbases of spins along $x$ is given by
\begin{eqnarray}
|\psi^-_{\alpha}\rangle_{12} &=& \frac{1}{2}(\alpha -\beta)
(|x+\rangle_1|x+\rangle_2 - |x-\rangle_1|x-\rangle_2) \nonumber \\
&+& \frac{1}{2} (\alpha + \beta) (|x+\rangle_1|x-\rangle_2 -
|x-\rangle_1|x+\rangle_2).
\end{eqnarray}
Thus, the correlations for spin measurements along $x$ are not
perfect.

If quantum physics is to be viewed as a science about what we can
say about possible measurement results and if, as we just have
seen, we cannot assert definite truth values to all possible
propositions, then we suggest to introduce the notion of
information, or our knowledge about these truth values. It is then
natural to refer to the amount of information on the truth values
of propositions.

Elsewhere we introduced an appropriate measure of information to
quantify this knowledge in quantum
mechanics~\cite{bruknerzeilinger}. If we denote the probability
for spins of the particles to be equal along $x$ by $p^+_{xx}$ and
the probability for them to be opposite by $p^-_{xx}$, the measure
is given by
\begin{equation}
I_{xx}= (p^+_{xx} - p^-_{xx})^2. \label{measure}
\end{equation}
Similarly, information about the spins of two particles along $z$
directions is given by $I_{zz}= (p^+_{zz} - p^-_{zz})^2$. In the
case of the four partially entangled states
(\ref{partialbell1}-\ref{partialbell2}) we have $I_{zz}=1$ and
$I_{xx}=4 \alpha^2 \beta^2$ and for the maximally entangled states
$I_{xx}=I_{zz}=1$.

Suppose Victor subjects particles 1 and 2 to a measurement in a
generalized Bell-state analyser, and he finds them in the not
maximally entangled state $|\psi^-_\alpha\rangle_{12}$. Then,
according to Eq.~(\ref{entswap}) particles 0 and 3 will be in the
also not maximally entangled state $|\psi^-_\beta\rangle_{03}$
(which is defined by inserting $\beta$ instead of $\alpha$ in the
expression (\ref{bas}) for $|\psi^-_\alpha\rangle_{12}$). The
reason for this can be seen even without involving
Eq.~(\ref{entswap}) by logically combining propositions and using
our measure of information. Because we observe particles 1 and 2
in the state $|\psi^-_{\alpha}\rangle_{12}$ our knowledge on
possible outcomes for measurements of spins of particles 1 and 2
along directions $z$ and $x$ is represented by $I^{12}_{zz}=1$ and
$I^{12}_{xx}=4 \alpha^2 \beta^2$, respectively. We note again that
with this we mean our a priori information about what outcome will
occur if the corresponding experiment is performed. Initially we
had prepared the pair of particles 0 and 1 in the state
$|\psi^-\rangle_{01}$ and the pair 2 and 3 in the state
$|\psi^-\rangle_{23}$ which means that for the pair $0-1$ we have
$I^{01}_{zz}=I^{01}_{xx}= 1$ and similarly for the pair $2-3$ we
have $I^{23}_{zz}=I^{23}_{xx}= 1$. By combining propositions in a
logical chain, which in the case of incomplete knowledge of joint
truth values corresponds to a multiplication of the corresponding
measures of information, we obtain
\begin{eqnarray}
I^{03}_{zz}&=&I^{01}_{zz} \cdot I^{12}_{zz} \cdot I^{23}_{zz}=1 \\
I^{03}_{xx}&=&I^{01}_{xx} \cdot I^{12}_{xx} \cdot I^{23}_{xx}=4
\alpha^2 \beta^2 \label{chain}
\end{eqnarray}
for the measures of information for the pair of particles 0 and 3
for measurements along $z$ and $x$, respectively. These values
indeed correspond to the state $|\psi^-_\beta\rangle_{03}$.
Interestingly, we obtained this by using only operations on
measures of information without involving the standard quantum
formalism.

In Ref.~\cite{essence} we suggested to describe the total
information contained in correlations of two particles as the sum
over the individual measures of information $I_{zz}$ and $I_{xx}$
carried in a set of spin measurements of the two particles along
$z$ and $x$ directions, i.e.,
\begin{equation}
I_{corr} \equiv I_{zz}+ I_{xx}. \label{corr}
\end{equation}
Here the value of $I_{corr}$ is maximized over all possible
choices of local coordinate systems (i.e. local $z$ and $x$
directions) for particles 1 and 2. The measurements of spin
products along $z$ and along $x$ are \emph{mutually complementary
for product states}. That is, for any product state (and, more
general, for any separable state) complete knowledge contained in
one of the observations in Eq.~(\ref{corr}) excludes any knowledge
in the other observation. For example, for the case of the product
state $|z+\rangle_1 |z+\rangle_2$ one has full information about
spins along $z$ $(I_{zz}=1)$ at the expense of complete lack of
information about spins along $x$ $(I_{xx}=0)$. This suggests that
$I_{corr}$ can be considered as an information-theoretic measure
of entanglement in the spirit of Schr\"{o}dinger's ''entanglement
of our knowledge"~\cite{schroedinger}. The value of $I_{corr}$ is
smaller than or equal to 1 for separable states, whereas it is 2
for the maximally entangled state. In the intermediate case of a
partially entangled state $|\psi^-_{\alpha}\rangle_{03}$ this
value is between 1 and 2 bits of information, i.e.
\begin{equation}
I^{03}_{corr} = 1 + 4\alpha^2 \beta^2. \label{cor}
\end{equation}

Quantum entanglement displays one of the most distinct features of
quantum mechanics against the classical world - conflict with
local realism - as quantified by violation of Bell's
inequalities~\cite{bell}. An example of such an inequality is the
Clauser-Horne-Shimony-Holt~\cite{chsh} (CSHS) inequality
\begin{equation}
S=|E(\phi,\tilde{\phi}) + E(\phi,\tilde{\phi}') +
E(\phi',\tilde{\phi}) - E(\phi',\tilde{\phi}')| \leq 2,
\end{equation}
where $2$ on the right-hand side is the local realistic limit, $S$
is the ``Bell parameter'', and $E(\phi,\tilde{\phi})$ is the
correlation function for measurements of two particles with local
measurement setting $\phi$ for the first particle and
$\tilde{\phi}$ for the second particle, respectively.

In Ref.~\cite{essence} we showed that the condition that
\emph{more than one bit} is contained in correlations of
measurements on a pair of particles is a \emph{necessary} and
\emph{sufficient} condition~\cite{horodecki} for violation of the
CHSH inequality by the pair of particles. Mathematically, we have
\begin{equation}
I_{corr} > 1 \Leftrightarrow S>2.
\end{equation}
It can be shown that the relation between the Bell parameter and
the amount of information contained in correlations is given by
$I_{corr}= S^2/4$ \cite{essence}.

In Ref.~\cite{pan} entanglement swapping was first experimentally
demonstrated, but the low photon-pair visibility prevented a
violation of a Bell's inequality for photons 0 and 3, which is an
important test to confirm the quantum nature of teleportation.
This is because the violation of Bell's inequalities for the
measurements on photons 0 and 3 indicates that no significant
information about the state of teleported photon 1 was gained in
the teleportation procedure and, consequently, the teleportation
is of high fidelity. This again can be seen from our
information-theoretical considerations. Suppose that Victor finds
particles 1 and 2 in the state $|\psi^-_\alpha\rangle_{12}$. Then
we know that particle 1, if it is measured along $z$, will give
result $+1$ with probability $\beta^2$ and result $-1$ with
probability $\alpha^2$. This knowledge about the individual
properties of particle 1 can be defined as given by (this is
analogue of definition (\ref{measure}); see
also~\cite{bruknerzeilinger})
\begin{equation}
I^{1}_{ind} = (p^+_z - p^-_z)^2 = (\beta^2-\alpha^2)^2
\label{ind},
\end{equation}
where, e.g., $p^+_z$ is the probability to observe result $+1$ for
measurement of spin of particle 1 along $z$.

We now give a ''complementarity relation" for the amount of
information contained in individual properties of particle 1 and
the amount information contained in correlations, or joint
properties, of the pair of particles 0 and 3. By summing Eq.
(\ref{cor}) and (\ref{ind}) we obtain
\begin{equation}
I^1_{ind} + I^{03}_{corr} = (\beta^2-\alpha^2)^2 + 1+ 4 \alpha^2
\beta^2 =2. \label{complem}
\end{equation}
Note that for symmetry reasons this is also true with particle 2
instead of 1. Equation~(\ref{complem}) shows that if the pair of
particles 0 and 3 violate the CHSH inequality, that is, if the
particles carry more than one bit of information in their joint
properties ($I^{03}_{corr}>1$), then information about individual
properties of particle 1 will be incomplete ($I^1_{ind}<1$), and
vice versa. In the extreme case of maximal violation of the
inequality no information is available to define properties of
particle 1. Consequently, the teleportation state is of perfect
fidelity.

One can conjecture that the complementary relation in the form of
the inequality $I^1_{ind} + I^{03}_{corr} \leq 2$ is also valid
for more general type of measurements (see also~\cite{banaszek}).
Then the inequality sign can be understood as coming from the
possibility that information can flow out of the systems into the
environment.

If Alice and Bob can only communicate via classical channel, they
cannot transfer an arbitrary quantum state with a fidelity of one
(here we assume that Alice has also access to Victor's measurement
station). Instead, on the basis of her measurement result Alice
can make a state estimation of the input state and send this
information to Bob who can reconstruct the state on the basis of
her result. The fidelity $f$ of this ``classical teleportation" is
then defined as the overlap between the input state
$|\psi_{in}\rangle$ and the state estimated $\rho_{est}$, i.e.
$f=\langle \psi_{in}| \rho_{est}| \psi_{in}\rangle$.

Somewhat lengthy but otherwise straightforward calculation shows
that fidelity of state estimation of particle 1, that is based on
the measurement defined by eigenstates
(\ref{partialbell1}-\ref{partialbell2}) is given by $f=
\frac{2}{3} (1-\alpha^2\beta^2)= \frac{1}{2} +
\frac{I^1_{ind}}{6}$. Here $f$ is averaged over all possible input
states of particle 1. Note that while $f$ refers to our knowledge
about the pre-measurement state of particle 1, $I^1_{ind}$
corresponds to our knowledge about the post-measurement properties
of this particle. Thus, if the state of particle 1 after the
measurement is completely undefined $(I^1_{ind}=0)$, the fidelity
is $f=1/2$ just as for the random choice. If, in contrast, the
measurement projects particle 1 into a well-defined (pure) state
$(I^1_{ind}=1)$, the fidelity achieves the value of $f_{cl}=2/3$.
This is known as the classical limit for teleportation and is
equal to the minimum fidelity of teleportation such that it can
still be considered as genuine quantum.

The high-fidelity teleportation and violation of the CHSH
inequality for measurements on photons 0 and 3 was recently
observed in the entanglement swapping experiment \cite{tomas}. The
value for $S$ obtained was $S=2.421 \pm 0.091$ which clearly
violates the local realistic limit of 2 by 4.6 standard deviations
as measured by the statistical error. Such a strong violation of
the CSHS inequality excludes a principal possibility of the use of
Alice's measurement result to classically reconstruct the
teleported state at Bob's side with high fidelity (e.g. via some
"hidden mechanism"). Since the experimental value for $S$
corresponds to $I^{03}_{corr}=S^2/4=1.465$, the maximum amount of
information $I^1_{ind}$ about particle 1 that could have been
gained that way is bounded by the complementarity
relation~(\ref{complem}): $I^1_{ind} \leq 0.536$. This results in
a fidelity limit of $f=0.589$, which is clearly lower than the
classical limit of $f_{cl}=2/3=0.666$.

In conclusion, in Peres words: "if we attempt to attribute an
objective meaning to the quantum state of a single system, curious
paradoxes appear: quantum effects mimic not only instantaneous
action-at-a-distance but also, as seen here, influence of future
actions on past events, even after these events have been
irrevocably recorded". In contrast, there is never a paradox if
the quantum states is viewed as to be no more than a
representative of our information. Furthermore such a view
provides us with both conceptually and formally much simpler
approach. This we demonstrate here for the entanglement swapping
experiment by deriving a quantitative complementarity relation
between the measure of information about the input state for
teleportation and the amount of entanglement of the resulting
swapped entangled state.

\begin{acknowledgments}

This work has been supported by the Austrian Science Foundation
(FWF) Project No. F1506, and by the European Commission, Contract
No. IST-2001-38864 (RamboQ). M.A. has been supported by the
Alexander von Humboldt foundation. {\v C}.B. has been supported
by the European Commission, Marie Curie Fellowship, Project No.
500764.

\end{acknowledgments}

\end{document}